\documentclass[aps,prl,twocolumn,groupedaddress]{revtex4-1}

\usepackage[version=3]{mhchem}
\usepackage{amsmath,amssymb,graphicx,color}
\usepackage{dcolumn}
\usepackage{bm}
\usepackage[mathlines]{lineno}
\usepackage{epsfig}
\usepackage{ulem} 
\usepackage{graphicx}
\usepackage[cp1251]{inputenc}
\usepackage{epstopdf}
\usepackage[T1,T2A]{fontenc}
\newcolumntype{P}[1]{>{\centering\arraybackslash}m{#1}}

\graphicspath{{figures/}}

\begin{document}
	
\title{All-optical switching and unidirectional plasmon launching with electron-hole plasma driven silicon nanoantennas}
\author{Sergey~V.~Li$^{1}$, Alexander~E.~Krasnok$^{2}$, Sergey~Lepeshov$^{1}$, Roman~S.~Savelev$^1$, Denis~G.~Baranov$^3$, and Andrea~Al\'{u}$^{2}$}
\affiliation{$^1$ITMO University, St. Petersburg 197101, Russia}
\affiliation{$^2$Departement of Electrical and Computer Engineering, The University of Texas at Austin, Texas 78712,USA}
\affiliation{$^3$Department of Physics, Chalmers University of Technology, 412 96 Gothenburg, Sweden}
	
\begin{abstract}	
High-index dielectric nanoparticles have become a powerful platform for modern light science, enabling various fascinating applications, especially in nonlinear nanophotonics for which they enable special types of optical nonlinearity, such as electron-hole plasma photoexcitation, which are not inherent to plasmonic nanostructures. Here, we propose a novel geometry for highly tunable all-dielectric nanoantennas, consisting of a chain of silicon nanoparticles excited by an electric dipole source, which allows tuning their radiation properties via electron-hole plasma photoexcitation. We show that the slowly guided modes determining the Van Hove singularity of the nanoantenna are very sensitive to the nanoparticle permittivity, opening up the ability to utilize this effect for efficient all-optical modulation. We show that by pumping several boundary nanoparticles with relatively low intensities may cause dramatic variations in the nanoantenna radiation power patterns and Purcell factor. We also demonstrate that ultrafast pumping of the designed nanoantenna allows unidirectional launching of surface plasmon-polaritons, with interesting implications for modern nonlinear nanophotonics.
\end{abstract}

\maketitle

\section{Introduction}

In the last several years high-index dielectric nanoparticles and nanostructures~\cite{Kuznetsov2016, krasnok2012all, jahani2016all} proved to be a promising platform for various nanophotonic applications, in particular for the design of functional nanoantennas~\cite{rolly2013controllable, krasnok2014superdirective, li2015all}, enhanced spontaneous emission~\cite{Regmi2016, krasnok2015enhanced, Bonod16, sun2016fluorescence}, photovoltaics~\cite{Brongersma2014}, frequency conversion~\cite{carletti2015enhanced, makarov2016self, Shorokhov2016}, Raman scattering~\cite{dmitriev2016resonant}, and sensing~\cite{Caldarola2015}. The great interest in such nanostructures is caused mainly by their ability to control the electric and magnetic components of light at the nanoscale~\cite{Kuznetsov2016}, while exhibiting low dissipative losses inherent to the materials with a negligible concentration of free charges~\cite{Caldarola2015}. In particular, it has been demonstrated that nanoantennas composed of high-index nanoparticles have the ability to realize directional scattering of the incident light and to effectively transform the near field of feeding quantum sources into propagating electromagnetic waves~\cite{krasnok2012all}.

Modification of the spontaneous emission rate of a quantum emitter induced by its environment, known as the Purcell effect~\cite{Purrel1946PR, Pelton}, is not so pronounced in all-dielectric structures~\cite{Bozhevolnyi2016, krasnok2012all}, as in microcavities~\cite{Vahala_2003} or plasmonic nanoantennas~\cite{Russell2012, Akselrod2014}. This is due to the fact that optical resonances of high-index nanoparticles are characterized by relatively low quality factors and large mode volumes, which results in low efficiency of light-matter interaction. However, it was recently shown that this disadvantage can be overcome by relying on the Van Hove singularity of a chain of high-index nanoparticles~\cite{krasnok2016demonstration}. Such approach allows to substantially enhance the local density of optical states (LDOS) at the location of a quantum source and thus achieve high values of the Purcell factor with relatively small dielectric nanostructures while leaving unaltered all their other advantages.

\begin{figure}[!b]
		\includegraphics[width=0.5\textwidth]{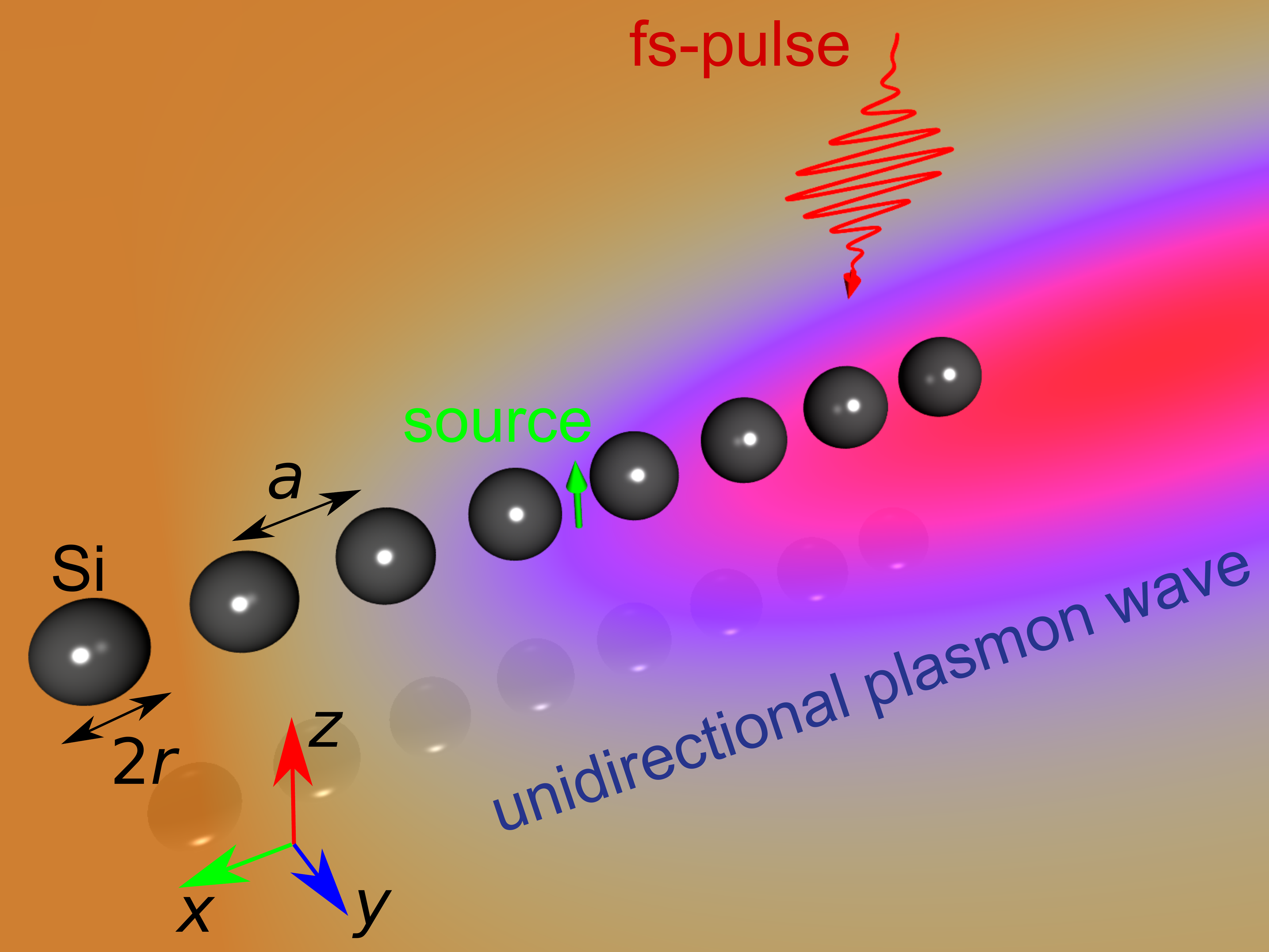}
		\caption{The schematic presentation of all-dielectric nanoantenna driven by electron-hole plasma photoexcitation via fs-laser pulse pumping of few boundary nanoparticles. The nanoantenna allows tuning the LDOS and radiation power pattern of a source (green arrow), and enables unidirectional launching of propagating plasmonic surface waves.}\label{concept}
\end{figure}

High-index dielectric nanostructures are also of a special interest for nonlinear nanophotonics, because they can exhibit strong nonlinear responses. It was recently predicted and experimentally demonstrated that photoexcitation of dense electron-hole plasma (EHP) in silicon (Si) nanoparticles~\cite{makarov2015tuning, Shcherbakov2015, Baranov2016} and nanodimers~\cite{baranov2016tuning} by femtosecond laser (fs-laser) pulses is accompanied by a dramatic modification of the radiation properties, whereas generation of EHP in germanium nanoantennas can even turn them into plasmonic ones in the mid-IR region~\cite{GePRL}. Here, we propose a highly tunable all-dielectric nanoantenna, consisting of a chain of Si nanoparticles excited by an electric dipole source, which allows for tuning its radiation properties via electron-hole plasma photoexcitation. We theoretically and numerically demonstrate the tuning of radiation power patterns and Purcell factor by pumping several boundary nanoparticles in the chain with relatively low peak intensities of fs-laser pulses. Moreover, we show that the proposed nanoantenna, being driven by fs-laser pulses, allows unidirectional launching of surface plasmon waves (Fig.~\ref{concept}), making this solution attractive for all-optical light manipulation systems.

\section{Results and discussion}

To briefly recall the origin of the \textit{Van Hove singularity}, let us consider a general periodic one-dimensional system supporting a set of guided modes. Decomposing its Green tensor into a series of eigenmodes, one can calculate the Purcell factor in such a system according to Ref.~\cite{Hughes}
\begin{equation}
{F} \simeq \frac{1}{\pi }{\left( {\frac{\lambda }{2}} \right)^2}\frac{c}{{{A_{{\rm{eff}}}}{V_{\rm gr}}}},
\end{equation}
with $A_{\rm eff}$ being the effective area of the resonant guided mode, $\lambda$ the free space wavelength, $V_{\rm gr}$ the group velocity of the mode, and $c$ is the speed of light. The divergence of the Purcell factor, occurring at the point of zero group velocity, is known as a Van Hove singularity. This expression clearly suggests that the Purcell factor benefits from slow light modes of the structure. In reality, its finite size prevents this divergence, but nevertheless largely enhanced Purcell factor can still be traced to the Van Hove singularity of the original structure.

\begin{figure}[!t]
	\includegraphics[width=0.49\textwidth]{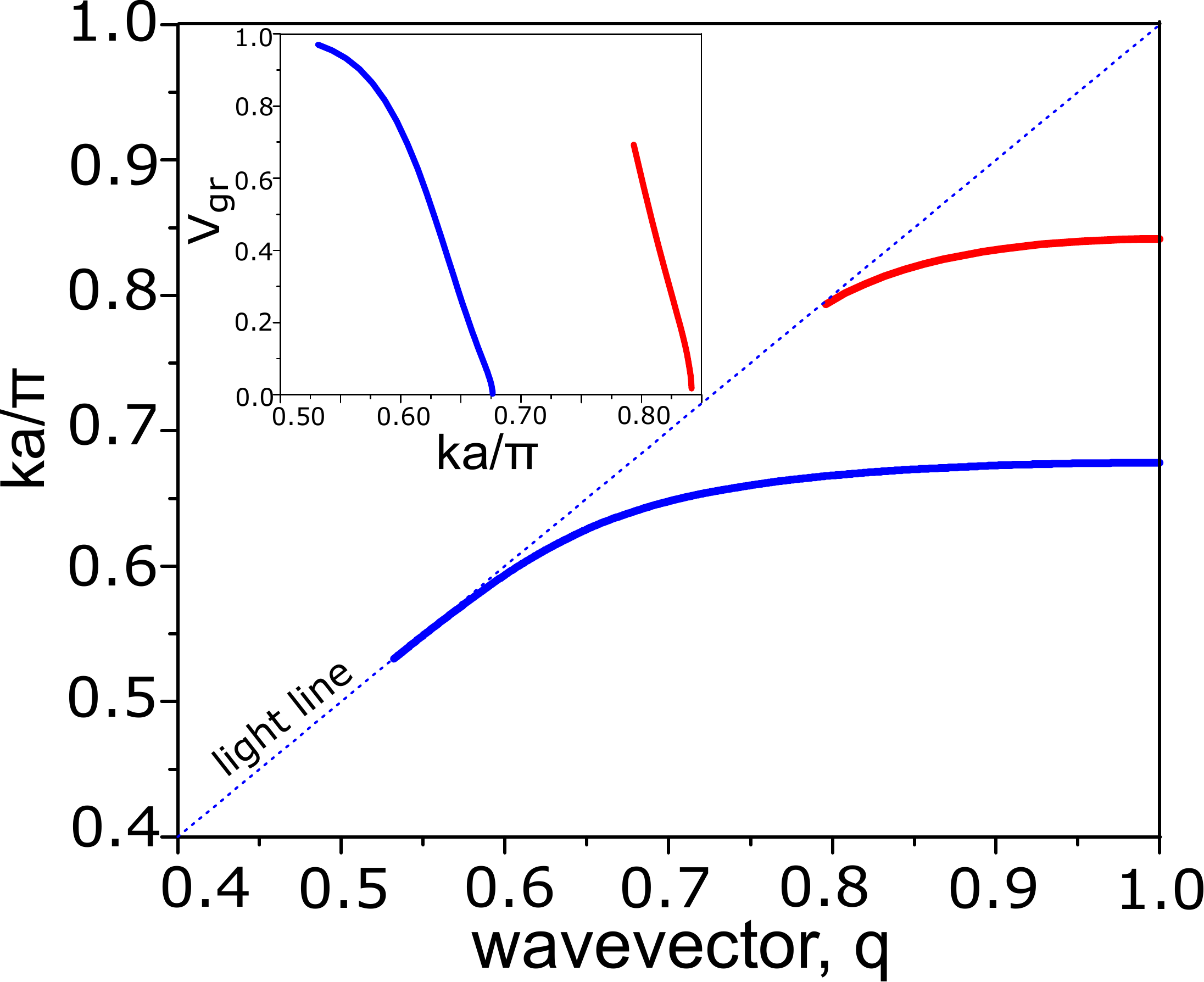}
	\caption{Dispersion curves of the eigenmodes in an infinite dielectric chain. Inset: dispersion curves of group velocity ($V_{\rm gr}$) of waveguiding modes in the infinite dielectric chain. The blue and red curves correspond to TM- and TE-modes, respectively.}\label{dispers}
\end{figure}

Here, we realize a Van Hove singularity using a chain of $N$ spherical dielectric nanoparticles excited by an electric dipole (green arrow), placed in the center of the chain and perpendicularly oriented to the chain axis, Fig.~\ref{concept}. We choose Si particles with dielectric permittivity $\varepsilon_1$ close to 16 in the operational frequency range~\cite{Aspnes1983}. The nanoparticles have all the same radius $r$ and the center-to-center distance between neighboring particles is $a$. 

The optical properties of nanoparticle-based nanoantennas can be understood from the infinite chain modal dispersion~\cite{Koenderink_09_NLetters, Alu_2006}. Therefore we start by calculating the optical properties of the infinite structure with an analytical approach, based on the well-known coupled-dipole model. Each particle is modeled as a combination of magnetic and electric dipoles with magnetic $\textbf{m}$ and electric $\textbf{p}$ momenta, oscillating with frequency $\omega$ [$\propto \exp(-i\omega t)$]. In the CGS system this approach leads to the linear system of equations:
\begin{align}\label{eq1}
\mathbf{p}_i &={\alpha_e}_i \sum\limits_{j \ne i}\left( \widehat{C}_{ij} \mathbf{p}_j - \widehat{G}_{ij} \mathbf{m}_j \right),\\ \nonumber
\mathbf{m}_i &={\alpha_m}_i \sum\limits_{j \ne i} \left( \widehat{C}_{ij} \mathbf{m}_j + \widehat{G}_{ij} \mathbf{p}_j \right),
\end{align}
where $\widehat{C}_{ij} = A_{ij}\widehat{I} + B_{ij}(\widehat{\mathbf{r}}_{ij} \otimes \widehat{\mathbf{r}}_{ij})$, $\widehat{G}_{ij} = - D_{ij}\widehat{\mathbf{r}}_{ij} \times \widehat{I}$, $\otimes$ is the dyadic product, $\widehat{I}$ is the unit $3 \times 3$ tensor, $\widehat{\mathbf{r}}_{ij}$ is the unit vector in the direction from $i$-th to $j$-th sphere, and
\begin{align}
A_{ij} &=\dfrac{\exp(i k_h r_{ij})}{r_{ij}} \left( k_h^2-\dfrac{1}{r_{ij}^2}+\dfrac{i k_h}{r_{ij}} \right),\\ \nonumber
B_{ij} &=\dfrac{\exp(i k_h r_{ij})}{r_{ij}} \left( -k_h^2 + \dfrac{3}{r_{ij}^2} - \dfrac{3 i k_h}{r_{ij}} \right),\\ \nonumber
D_{ij} &=\dfrac{\exp(i k_h r_{ij})}{r_{ij}} \left( k_h^2 + \dfrac{i k_h}{r_{ij}} \right),
\end{align}
where $r_{ij}$ is the distance between the centers of $i$-th and $j$-th spheres, $\varepsilon_h $ is the permittivity of the host medium, $k_h=\sqrt{\varepsilon_h}\omega/c$ is the host wavenumber, $\omega=2\pi\nu$, and $\nu$ is the frequency. The quantities $\alpha_m$ and $\alpha_e$ are the magnetic and electric polarizabilities of a spherical particle~\cite{Bohren}:
\begin{equation}\label{eq2}
\alpha_e=i\dfrac{3\varepsilon_h a_1}{2k^3_h}, ~~~\alpha_m=i\dfrac{3 b_1}{2k^3_h},
\end{equation}
where $a_1$ and $b_1$ are electric and magnetic Mie coefficients. The coupled dipole approximation outlined above is justified for the geometrical parameters of the nanoparticles and their relative distance~\cite{Savelev2014}.

The solution of Eq.~\ref{eq1} without source [dispersion of waveguide eigenmodes $\omega(k)$] for the infinite dielectric chain with $r=70$~nm and $a=200$~nm in free space is shown in Fig.~\ref{dispers}. Here, we use the dimensionless wavenumber $q = \beta a/\pi$, where $\beta$ is the Bloch propagation constant. The blue and red curves correspond to transverse magnetic (TM) and transverse electric (TE) modes, respectively, which are the only modes excited by the dipole source with the chosen orientation. Both of these modes are characterized by induced magnetic and electric moments (except the points at the band edge). Due to the spectral separation of resonances of the single particle, the magnetic moments are dominant in the first branch (TM), and electric moments in the second one (TE). The inset in Fig.~\ref{dispers} shows the calculated group velocities of the waveguide modes as a function of frequency. It can be seen that the group velocity $V_{\rm gr}$ drops to zero at the band edge around $ka/\pi \approx 0.675$ and $ka/\pi \approx 0.83$. Since the symmetry of the ED source matches the symmetry of the TM staggered mode (and not the TE one), we may expect significant enhancement of the Purcell factor for a finite system around the first frequency.

\begin{figure}[!t]
	\includegraphics[width=0.5\textwidth]{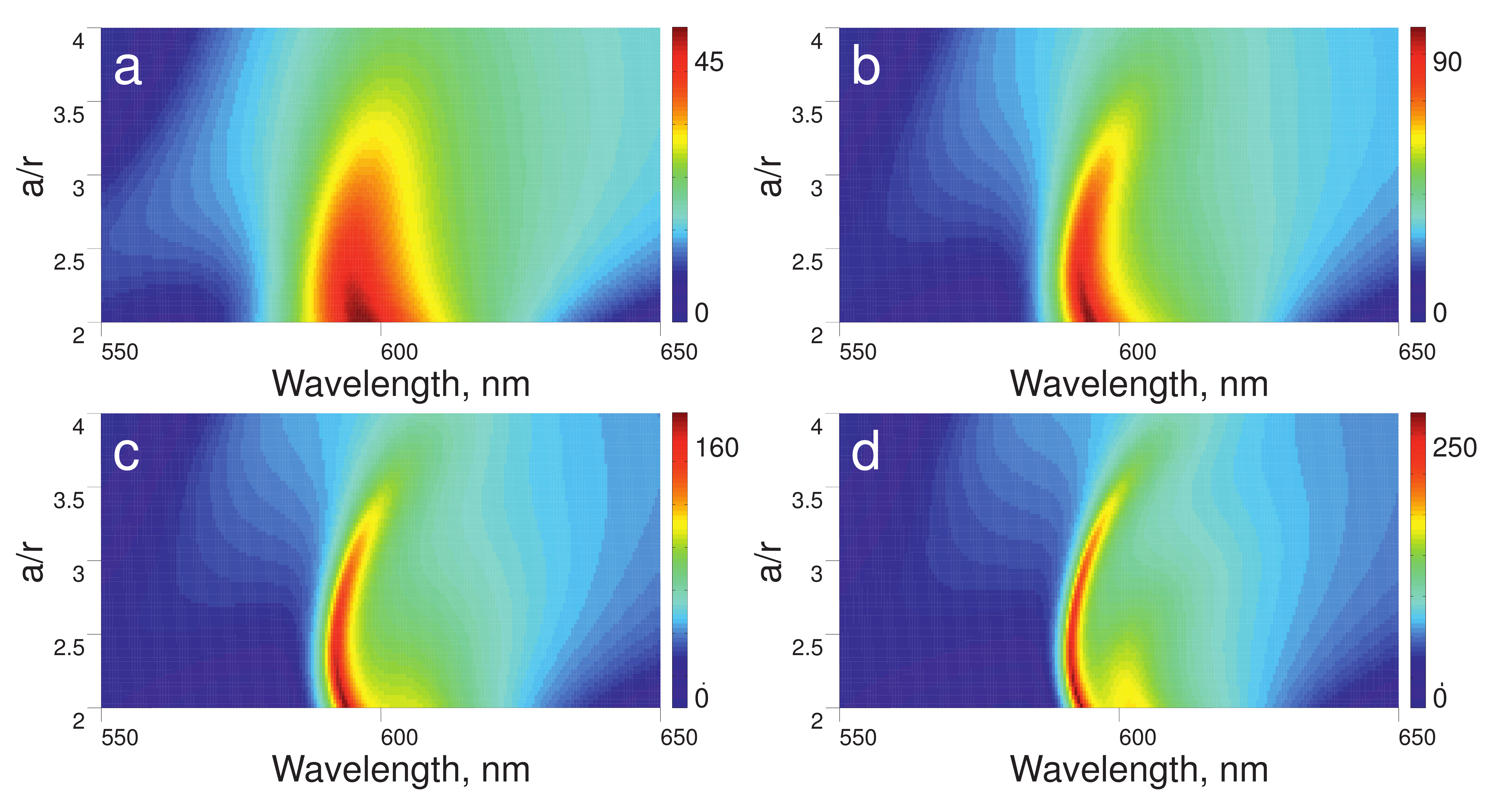}
	\caption{The log-scale Purcell factor as the function of  the radiation wavelength and ratio $r/a$ for the dielectric chain with $\varepsilon_1=16$, for different number of nanoparticles: (a)~$N=4$, (b)~$N=6$, (c)~$N=8$, and (d)~$N=10$; $r$ is taken equal to 70~nm, $a$ is a period of the chain.}\label{2fig}
\end{figure}

To confirm this expectation, we calculate the Purcell factor using the Green's tensor approach~\cite{Novotny_Hecht_book}: 
\begin{equation}\label{eq3}
F = \frac{3}{{2k_h^3}}{\bf{z}} \cdot {\rm{Im}}[{\bf{G}}\left( {0,0;\omega } \right) ]\cdot {\bf{z}}
\end{equation}
with ${\bf{G}}\left( {0,0;\omega } \right)$ being Green's tensor of an electric dipole in the center of chain (point of the dipole source localization) and ${\bf{z}}$ being the unit vector pointing in the $z$ direction (Fig.~\ref{concept}). Fig.~\ref{2fig} shows the calculated Purcell factor as a function of wavelength and ratio $r/a$ for a dipole source located in the center of a dielectric chain with different number of particles: (a)~$N=4$, (b)~$N=6$, (c)~$N=8$, and (d)~$N=10$. We observe that increasing the number of nanoparticles $N$ gives rise to enhancement of Purcell factor. For example, the maximal value of the Purcell factor for $N=10$ reaches 250. Along with the calculations shown in Fig.~\ref{dispers}, we conclude that the maximum of Purcell factor arises around the Van Hove singularity.

Now, we are ready to show that the excitation of slow guided modes determining a Van Hove singularity is very sensitive to the electrodynamical properties of the system. Our aim is to utilize this effect to engineer highly tunable nanoantennas, for which relatively low intensities of external laser pulses can control and cause a dramatic modification of the optical properties of the material, and consequently the radiation properties (intensities of emission and power patterns) of the nanoantenna. To enable the switching of the nanoantenna properties we employ the nonlinear response caused by \textit{electron-hole plasma photoexcitation} the in boundary particles of the Si nanoantenna.

To describe EHP-induced tuning of the nanoantenna, we employ the analytical approach developed in Ref.~\cite{Baranov2016}. The dynamics of volume-averaged EHP density $\rho_{eh}$ is modeled via the rate equation
\begin{equation}\label{plasma}
\frac{{d{\rho _{{\text{eh}}}}}}{{dt}} = - \Gamma {\rho _{{\text{eh}}}} + \frac{{{W_1}}}
{{\hbar \omega }} + \frac{{{W_2}}}{{2\hbar \omega }},
\end{equation}
where, $W_{1,2}$ are the volume-averaged absorption rates due to one- and two-photon processes, and $\Gamma$ is the EHP recombination rate which depends on EHP density~\cite{Cardona}. The absorption rates are written in the usual form as ${W_1} = \frac{\omega }{{8\pi }} \left\langle {{{\left| {{\mathbf{\tilde E_{\rm in}}}} \right|}^2}} \right\rangle {\rm Im} (\varepsilon)$
 and ${W_2} = \frac{\omega }{{8\pi }} \left\langle {{{\left| {{\mathbf{\tilde E_{\rm in}}}} \right|}^4}} \right\rangle \operatorname{Im} {\chi ^{(3)}}$, where angle brackets denote averaging over the nanoparticle volume, and $\operatorname{Im} {\chi ^{(3)}} = \frac{{\varepsilon {c^2}}} {{8\pi \omega }}\beta $ with $\beta$ being two-photon absorption coefficient. The relaxation rate of EHP in c-Si is dominated by Auger recombination $\Gamma = \Gamma _{\text{A}} \rho_{\rm eh}^2$ with $\Gamma_{\rm A}=4 \cdot 10^{ - 31}$~s$^{-1}$cm$^6$ (Ref.~\cite{Shank}). Now the permittivity of photoexcited Si should be related to time-dependent EHP density:
\begin{equation}
\varepsilon( {\omega ,{\rho _{\rm eh}}}) = \varepsilon_0+\Delta\varepsilon_{\rm bgr}+\Delta\varepsilon _{\rm bf} + \Delta\varepsilon _{\rm D},
\label{eps}
\end{equation}
where ${\varepsilon _{{\text{0}}}}$ is the permittivity of non-excited material, while $\Delta\varepsilon_{\text{bgr}}$, ${\Delta\varepsilon _{{\text{bf}}}}$, and $\Delta\varepsilon _{{\text{D}}}$ are the contributions from bandgap renormalization, band filling, and Drude term, respectively. The detailed expressions for all contributions in Eq.~(\ref{eps}) can be found in Ref.~\cite{Baranov2016}. In total, these three contributions lead to decrease of the real part of permittivity with increasing EHP density.

\begin{figure}[!t]
	\includegraphics[width=0.5\textwidth]{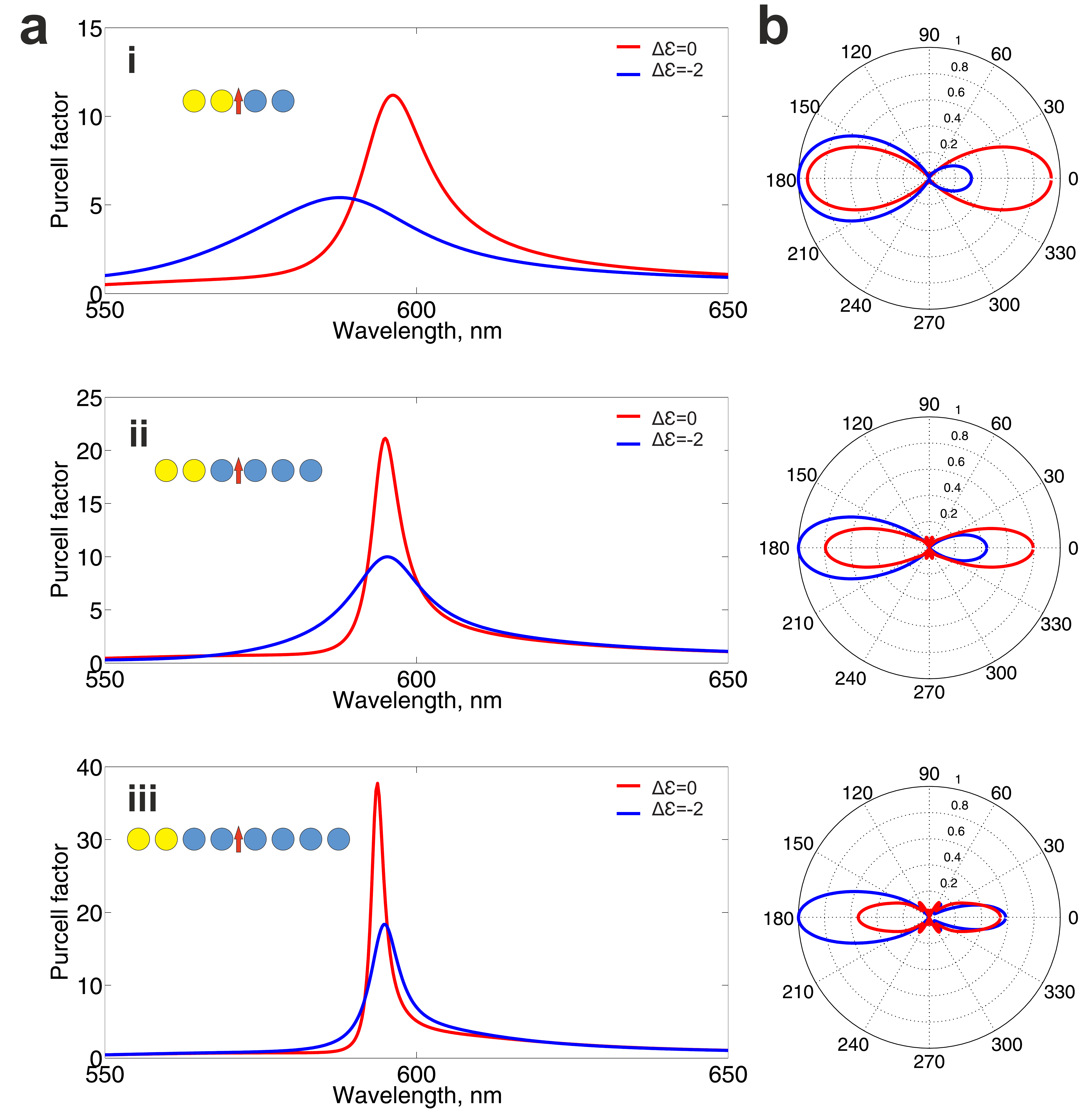}
	\caption{(a)~Spectral dependence of the Purcell factor for the chains of dielectric nanoparticles for different number of nanoparticles $N$: (i)~$N=4$; (ii)~$N=6$; (iii)~$N=8$. (b)~Radiation power patterns (E-plane) of the electric dipole source at the radiation wavelength of 600~nm. Red curves correspond to the unaffected chains ($\Delta\varepsilon=0$), whereas the blue ones correspond to the chains with photo-excited boundary particles ($\Delta\varepsilon=2$). }\label{3fig}
\end{figure}

The spectral dependency of the Purcell factor before and after EHP photoexcitation are presented in Fig.~\ref{3fig}(a) for a different number of particles $N$. The calculations are performed for the change of the real part of Si permittivity $\Delta \varepsilon=-2$ [where $\Delta \varepsilon=\varepsilon( {\omega ,{\rho _{\rm eh}}}) - \varepsilon_0$], which is achieved at the wavelength of 600~nm upon excitation of EHP with density $\rho_{eh}\approx 1.5 \cdot 10^{21}$ cm$^{-3}$. The decrease of the Purcell factor approximately by a factor of 2 in all cases (i--iii), along with the spectral broadening, are caused by symmetry breaking of the chain and corresponding decrease of the quality factor of the Van Hove singularity mode. The EHP photoexcitation also modifies the radiation pattern of the nanoantenna, Fig.~\ref{3fig}(b). Before the plasma excitation ($\Delta\varepsilon=0$) the radiation pattern has two symmetric lobes directed along the chain axis in forward and backward directions (red curves). In this case, the maximal value of directivity grows with increasing of $N$. After the plasma excitation ($\Delta\varepsilon=-2$), nanoantenna radiates mostly in the direction of the affected particles. We note that the degree of modification of the radiation pattern grows with increasing number of particles. For example, in the case of $N$=8~[Fig.~\ref{3fig}(b)iii], the directivity in the left direction is two times larger that in the right one. Thus, the EHP photoexcitation can be applied for all-optical switching of radiation patterns. Such dramatic tuning of the radiation pattern is caused by the Van Hove singularity regime of the initially unaffected nanoantenna.

\begin{figure}[!t]
	\includegraphics[width=0.49\textwidth]{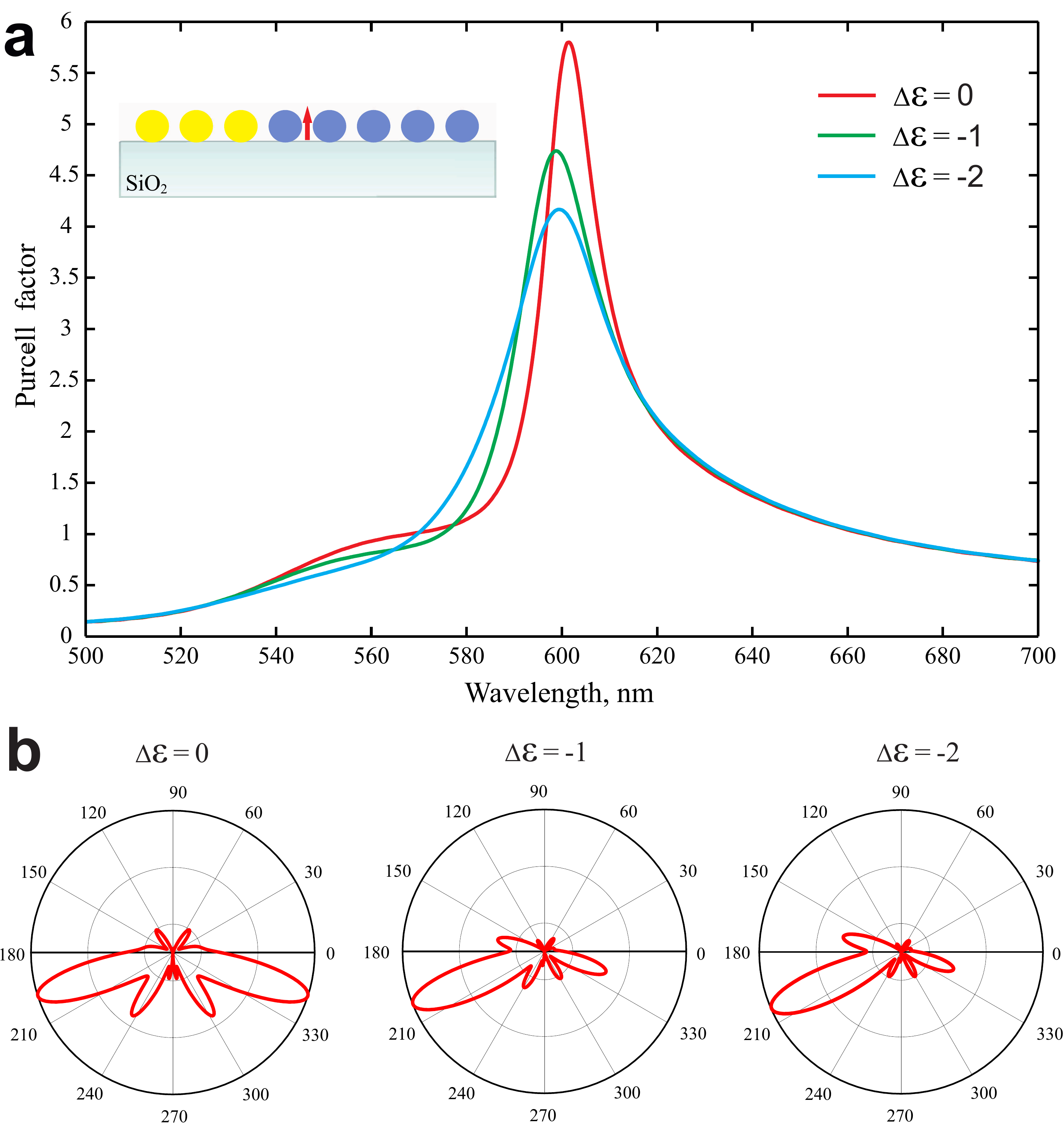}
	\caption{(a)~Purcell factor for the chain of $N=8$ dielectric nanoparticles arranged on a SiO$_2$ substrate as a function of radiation wavelength. Difference of the dielectric permittivities of unaffected and affected particles are $\Delta\varepsilon=0$ (red curve), $\Delta\varepsilon=-1$ (green curve), and $\Delta\varepsilon=-2$ (blue curve). (b)~Power patterns for different $\Delta\varepsilon$.}\label{4fig}
\end{figure}

In the vast majority of nanoantenna realizations, the substrate substantially affects the nanoantenna characteristics (see, for example, Ref.~\cite{Krasnok2015}). For this reason, we analyze how a SiO$_2$ substrate affects the nanoantenna's characteristics. To simulate the nanoantenna consisting of 8 nanoparticles, located on the  SiO$_2$ substrate with $\varepsilon_{\rm sub}=2.21$, we utilize the commercial software CST Microwave Studio. To calculate the Purcell factor, the method based on the input impedance of a small (in terms of radiation wavelength) dipole antenna~\cite{krasnok2015antenna} has been applied. The corresponding results are presented in Fig.~\ref{4fig}(a). These spectra qualitatively agree with our analytical calculations presented above. However, in this case, the substrate breaks the mirror symmetry with respect to the $z$ axis, which leads to a reduction of Purcell factor. Moreover, the numerical calculations reveal that the sharpest effect of EHP photoexcitation on the radiation pattern manifests itself when three boundary particles are illuminated [see inset in the Fig.~\ref{4fig}(a)]. During the EHP photoexcitation, the maximum of Purcell factor slightly decreases from 5.9 to 4.7 for $\Delta\varepsilon=-1$ and to 4.1 for $\Delta\varepsilon=-2$ accompanied by shifting of the resonant frequency to the shorter wavelengths due to the decrease of boundary particles dielectric permittivity.

\begin{figure}[t]
\includegraphics[width=0.5\textwidth]{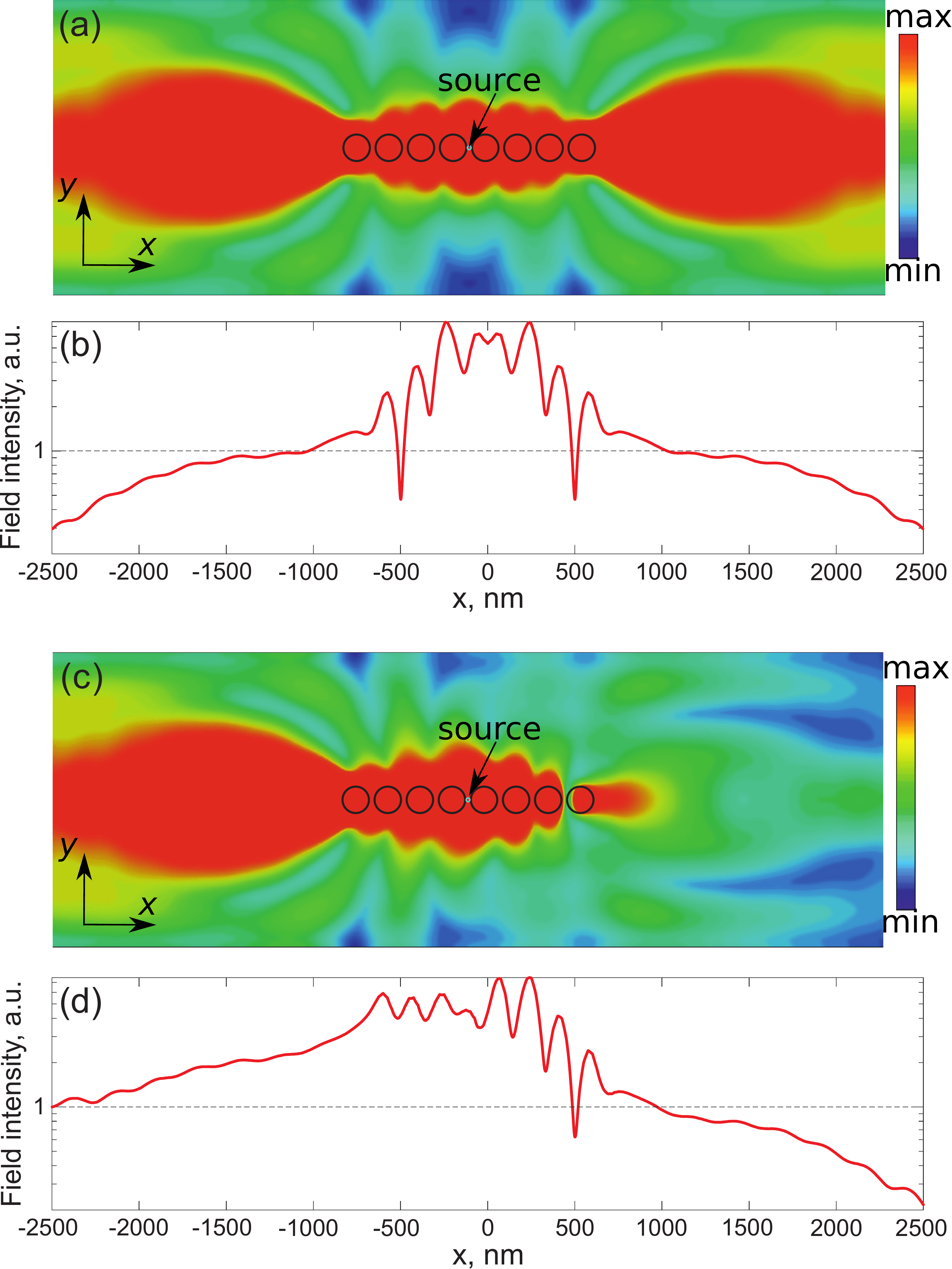}
\caption{Radiation of the nanoantenna composed by 8 Si nanoparticles placed with period 200~nm on a silver substrate with the 60~nm spacer glass layer; the dipole source is located at the center of the chain perpendicularly to the substrate. (a,c)~Electric field distribution profiles in the plane orthogonal to the source dipole in the cases of unaffected and affected three boundary nanoparticles, respectively. (b,d)~Electric field intensities as a function of $x$ coordinate in the cases of unaffected and affected three boundary nanoparticles, respectively.}
\label{fig6}
\end{figure}

Fig.~\ref{4fig}(b) demonstrates the change in radiation pattern induced by EHP photoexcitation for the nanoantenna on SiO$_2$ substrate. When $\Delta\varepsilon$=0 (unaffected nanoantenna) the radiation power pattern is symmetric with respect to the dipole axis and has two main lobes. It can be observed that the mirror symmetry of the radiation patterns with respect to the $z$ axis is broken, and the main lobes are oriented into the substrate, which has higher refractive index than the upper space. The modification of three boundary particles dramatically changes the power pattern: the reconfiguration is sufficient for practical applications even for $\Delta\varepsilon=-1$.

This effect can be used for the unidirectional launching of waveguide modes in plasmonic waveguides at will. Fig.~\ref{fig6} demonstrates this phenomenon, showing the calculated radiation of a nanoantenna composed of 8 Si nanoparticles placed with the period of 200~nm on a silver substrate with the 60~nm spacer glass (SiO$_2$) layer. The SiO$_2$ spacer serves as a buffer layer for the protection of silver substrate from sulfidation. The dipole source is located at the center of chain perpendicular to the substrate. Figs.~\ref{fig6}(a) and (b) show the electric field distribution profile and the electric field intensity as a function of the $x$ coordinate in the cases of the unaffected nanoantenna, respectively. It can be seen that the nanoantenna launches surface plasmons symmetrically to the positive and negative directions of axis $x$. However, when three boundary nanoparticles are illuminated by the pump beam, the nanoantenna launches surface plasmons almost unidirectionally, Figs.~\ref{fig6}(c,d). We achieve a value of front-to-back ratio up to 5 for this geometry, Fig.~\ref{fig6}(d).

As a final step of our analysis, we estimate the parameters of a pump pulse required for generation of $1.5 \cdot 10^{21}$ cm$^{-3}$ EHP, assumed in the electromagnetic calculations above. At the high intensities required for photoexcitation of Si, two-photon absorption (TPA) usually dominates over one-photon process~\cite{Baranov2016}. Silicon has a particularly large TPA coefficient between 600 and 700 nm~\cite{Reitze}. We set the wavelength of pump pulse to 650 nm in order to avoid the interference between pump and dipole source signals. Estimating the enhancement factors for $ \left\langle {{{\left| {{\mathbf{\tilde E_{\rm in}}}} \right|}^2}} \right\rangle$ and $ \left\langle {{{\left| {{\mathbf{\tilde E_{\rm in}}}} \right|}^4}} \right\rangle$ for a Si nanoparticle on a substrate and using equations (\ref{plasma}) to calculate the dynamics of EHP density, we find that a 200 fs pulse with peak intensity of 35~GW/cm$^2$ and 25~GW/cm$^2$ provides $1.5 \cdot 10^{21}$ cm$^{-3}$ EHP in the nanoparticle on a glass and silver substrates, respectively, during $\approx 1$ ps that should be sufficient for obtaining the degree of tuning demonstrated above.

\section{Conclusion}

In conclusion, we have proposed highly tunable all-dielectric nanoantennas, consisting of a chain of Si nanoparticles excited by an electric dipole source, that allow tuning their radiation properties via electron-hole plasma photoexcitation. We have theoretically and numerically demonstrated the tuning of radiation power patterns and the Purcell effect by additional pumping several boundary nanoparticles with relatively low peak intensities. We have also demonstrated that these effects remain valid for the nanoantenna situated on a dielectric surface. The proposed nanoantenna, driven by fs-laser pulses, also allows tunable unidirectional launching of surface plasmon waves, with interesting implications for modern nonlinear nanophotonics.

\section*{Acknowledgments}

This work was financially supported by Russian Science Foundation (Grant 15-19-30023) and by Russian Foundation for Basic Research (Project~16-37-60076). This work was also partially supported by the Air Force Office of Scientific Research.


%

\end{document}